\documentclass[runningheads]{llncs}

\usepackage[T1]{fontenc}
\usepackage{lmodern}

\usepackage{graphicx}
\usepackage{booktabs}
\usepackage{amsmath}
\usepackage{url}
\usepackage{flafter}
\usepackage{placeins}
\usepackage{tikz}
\usepackage{multirow}
\usetikzlibrary{arrows.meta,positioning}

\urldef{\codeurl}\url{https://github.com/HeeSeung-Shin/Brats2026-Goat-HS}

\begin{document}
\title{AIstatLab at BraTS-GoAT 2026 Task 3: Quality-Controlled Consensus Pseudo-Labeling and Dense Expert Adaptation}
\titlerunning{AIstatLab at BraTS-GoAT 2026}

\author{
Heeseung Shin\inst{1}\orcidID{0009-0004-0430-9547}
\and
Changwon Lim\inst{2}\orcidID{0000-0002-3182-8420}
\thanks{Corresponding author.}
}
\authorrunning{H. Shin and C. Lim}
\institute{
Department of AI in Medicine, Chung-Ang University,
Seoul 06974, Republic of Korea\\
\email{hs970416@cau.ac.kr}
\and
Department of Statistics and Data Science, Chung-Ang University,
Seoul 06974, Republic of Korea\\
\email{clim@cau.ac.kr}
}
\maketitle

\begin{abstract}
BraTS-GoAT Task~3 evaluates brain-tumor segmentation across heterogeneous
entities using shared ET, TC, and WT regions. We present a 3D
ResidualEncoderUNet trained with quality-controlled consensus pseudo-labels,
a training-only nested-region objective, and a four-expert bottleneck adapter.
Five-fold ResEnc-M/L teachers generated predictions for 1,138 unlabeled cases;
fusion, hierarchy enforcement, and confidence-and-agreement screening retained
983 for training with 1,351 annotated cases. The final five-fold ResEnc-M
entry achieved mean DSC 0.834, mean NSD 0.510, and
$A_{\mathrm{calc}}=0.6720$ on the validation server. Component effects were
modest and region-dependent, with the clearest gains on ET. Source code is
available at \codeurl.

\end{abstract}

\section{Introduction}

Automated multimodal-MRI segmentation supports quantitative assessment and
treatment planning, but cross-cohort generalization remains difficult because
tumor biology, lesion burden, imaging protocols, and populations vary%
~\cite{menze2015brats}. BraTS-GoAT Task~3 combines adult glioma, meningioma,
brain metastasis, pediatric tumor, and Sub-Saharan African cohorts under
shared enhancing-tumor (ET), tumor-core (TC), and whole-tumor (WT) regions%
~\cite{brats2026cluster}.

We build on a ResidualEncoderUNet baseline%
~\cite{isensee2024revisited,isensee2021nnunet} and offline
pseudo-labeling designed to limit confirmation bias%
~\cite{lee2013pseudolabel,arazo2020confirmation}.
The final $K=4$ ResEnc-M system was evaluated as a five-fold ensemble on
the official validation server. Pseudo-label/QC and paired-adapter analyses
used the 271-case fold-0 validation split of the organizer-released annotated
training set. Expert-capacity screening used fold-0 validation server
results, auxiliary supervision used matched five-fold validation server
configurations, and the heterogeneous-ensemble comparison was descriptive.

\begin{samepage}
Our main contributions are:
\begin{itemize}
\item An offline regional-consensus pipeline for quality-controlled
pseudo-labeling. It fuses predictions from two five-fold teacher
ensembles (ResEnc-M and ResEnc-L), enforces the ET/TC/WT hierarchy,
and screens candidates using confidence, uncertainty, foreground
coverage, and inter-teacher agreement. It retained 983 of 1,138
unlabeled candidates for training with the 1,351-case annotated
release.

\item A mixed-supervision ResEnc-M student that combines a training-only
auxiliary ET/TC/WT head with an input-conditioned dense $K=4$ residual
adapter using soft routing~\cite{puigcerver2024softmoe}. The auxiliary
head is bypassed at inference, whereas the adapter remains active.

\end{itemize}
\end{samepage}

\section{Related Work}
\label{sec:related-work}

BraTS established nested ET/TC/WT glioma segmentation
\cite{menze2015brats,baid2021brats,bakas2017tcga}; subsequent tasks
covered meningioma, brain metastasis, pediatric tumors, and Sub-Saharan
African populations
\cite{labella2023menchallenge,labella2024meningioma,moawad2025mets,kazerooni2023peds,kazerooni2024peds,adewole2026africa}, which BraTS-GoAT
evaluates jointly \cite{brats2026cluster}. Representative 3D backbones
include residual-encoder nnU-Net, MedNeXt, and SegMamba
\cite{isensee2021nnunet,isensee2024revisited,roy2023mednext,xing2024segmamba}.

Pseudo-labeling can reinforce prediction errors
\cite{lee2013pseudolabel,arazo2020confirmation}; teacher averaging,
uncertainty control, and cross-pseudo supervision improve target stability
\cite{tarvainen2017meanteacher,yu2019uamt,chen2021cps,luo2022urpc}.
Deep supervision and overlapping-region objectives complement mutually
exclusive voxel labels \cite{lee2015deep,isensee2019nonewnet}, while soft
expert routing supports input-dependent feature adaptation
\cite{jacobs1991moe,shazeer2017sparsemoe,puigcerver2024softmoe}.
These ideas motivate our offline two-family consensus, training-only
regional head, and dense soft-gated adapter.

\section{Method}

\begin{figure}[t]
    \centering
    \includegraphics[width=.98\linewidth]{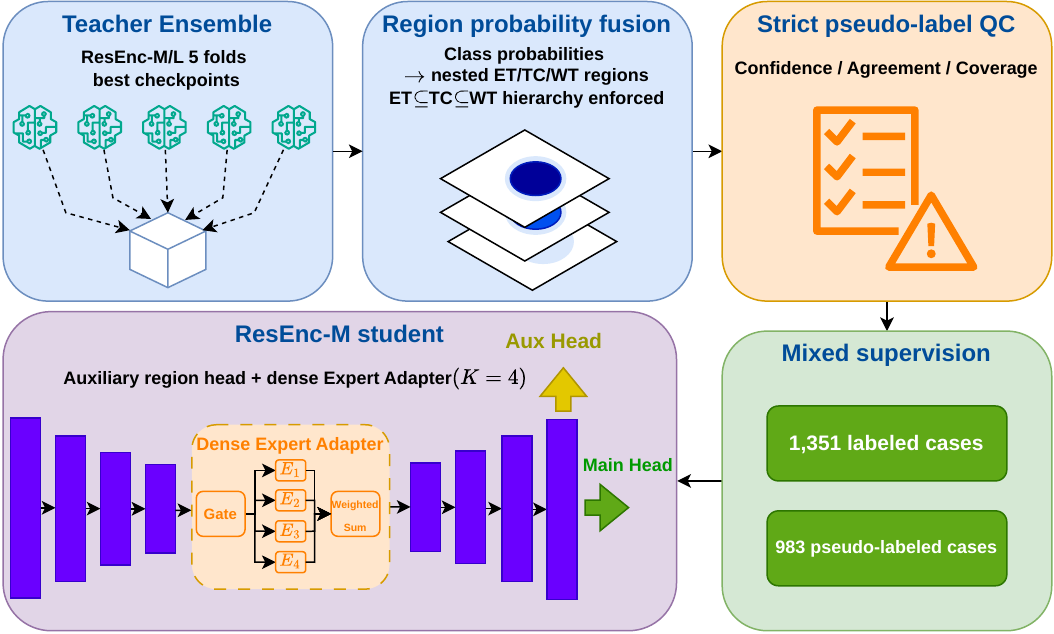}
    \caption{Overview of the proposed mixed-supervision framework.
Quality-controlled pseudo-labels from ResEnc-M and ResEnc-L teacher ensembles
augment labeled data for training a ResEnc-M student with a four-expert dense
adapter and an auxiliary region head.}
    \label{fig:framework}
\end{figure}

\subsection{Data and Consensus Pseudo-Label Generation}
\label{sec:data-pseudo-label}

\paragraph{Data and task representation.}
Each case contains T1-native (T1n), contrast-enhanced T1 (T1c), T2-weighted (T2w), and T2-FLAIR (T2f) MRI. BraTS-GoAT Task 3 combines multiple brain-tumor cohorts under four mutually exclusive classes: background (BG; 0), necrotic/non-enhancing tumor core (NCR/NET, abbreviated NCR; 1), peritumoral edema (ED; 2), and enhancing tumor (ET; 3). The evaluated nested regions are $\mathrm{ET}=\{3\}$, $\mathrm{TC}=\{1,3\}$, and $\mathrm{WT}=\{1,2,3\}$, with
$\mathcal{C}=\{\mathrm{BG},\mathrm{NCR},\mathrm{ED},\mathrm{ET}\}$ and $\mathcal{R}=\{\mathrm{ET},\mathrm{TC},\mathrm{WT}\}$.
According to the organizer description, the training release comprised the full BraTS adult-glioma training set with ground truth and meningioma and brain-metastasis training sets for which ground truth was available only for subsets~\cite{brats2026cluster}. The 1,351 annotated cases were divided into five folds (270--271 validation cases per fold), and the 1,138 volumes used for pseudo-labeling were provided as the organizer-released Task~3 unlabeled pool. Case-level cohort identity and clinical metadata were not released for this pool, precluding a direct cohort-wise breakdown of either the unlabeled pool or the strict-QC exclusions.
\paragraph{Teacher fusion and hierarchy enforcement.}
Five-fold ResidualEncoderUNet M and L teacher ensembles produced class probabilities $p_c^T(x)$ for voxel $x$, class $c\in\mathcal{C}$, and teacher family $T\in\{\mathrm{M},\mathrm{L}\}$. Within each family, logits from the five fold checkpoints were equally averaged and converted to probabilities. The class probabilities were converted to nested regions:
\begin{equation}
q_{\mathrm{ET}}^T=p_{\mathrm{ET}}^T,\qquad
q_{\mathrm{TC}}^T=p_{\mathrm{NCR}}^T+p_{\mathrm{ET}}^T,\qquad
q_{\mathrm{WT}}^T=p_{\mathrm{NCR}}^T+p_{\mathrm{ED}}^T+
p_{\mathrm{ET}}^T.
\label{eq:region-probabilities}
\end{equation}
For $r\in\mathcal{R}$, the teacher families were fused as
\begin{equation}
q_r=w_rq_r^{\mathrm{M}}+(1-w_r)q_r^{\mathrm{L}},\qquad
(w_{\mathrm{ET}},w_{\mathrm{TC}},w_{\mathrm{WT}})
=(0.70,0.50,0.40).
\end{equation}
After thresholding at 0.50 to obtain $\widetilde M_r$, the masks were
corrected as
\begin{equation}
M_{\mathrm{ET}}=\widetilde M_{\mathrm{ET}},\qquad
M_{\mathrm{TC}}=\widetilde M_{\mathrm{TC}}\cup M_{\mathrm{ET}},\qquad
M_{\mathrm{WT}}=\widetilde M_{\mathrm{WT}}\cup M_{\mathrm{TC}},
\end{equation}
ensuring $M_{\mathrm{ET}}\subseteq M_{\mathrm{TC}}\subseteq M_{\mathrm{WT}}$.
Before QC, 26-connected components smaller than 5, 3, and 1~mm$^3$ for WT, TC, and ET, respectively, were removed only when their mean and maximum probabilities were below 0.15 and 0.30. ET was additionally suppressed when its volume was below 1~mm$^3$ and its maximum probability was below 0.35.

\paragraph{Quality control.}
Candidates with an empty foreground region $F=M_{\mathrm{WT}}$ were
rejected. Let $\bar p_c=(p_c^{\mathrm{M}}+p_c^{\mathrm{L}})/2$, and let $\bar p_{(1)}$ and $\bar p_{(2)}$ be the largest and second-largest class probabilities. For nonempty $F$, we computed mean confidence $C_F$, class margin $M_F$, normalized entropy $H_F$, high-confidence fraction $Q_F$, and foreground coverage $\rho_F$:
\begin{equation}
\begin{aligned}
C_F&=\frac{1}{|F|}\sum_{x\in F}\bar p_{(1)}(x),&
M_F&=\frac{1}{|F|}\sum_{x\in F}
\bigl(\bar p_{(1)}(x)-\bar p_{(2)}(x)\bigr),\\
H_F&=-\frac{1}{|F|\log4}\sum_{x\in F}\sum_{c\in\mathcal{C}}
\bar p_c(x)\log\bar p_c(x),&\\
Q_F&=\frac{1}{|F|}\sum_{x\in F}
\mathbf{1}[\bar p_{(1)}(x)\geq0.75],\quad
\rho_F=\frac{|F|}{|\Omega|}.
\end{aligned}
\end{equation}
The confidence criterion was
\begin{equation}
I_{\mathrm{conf}}=
\mathbf{1}\!\left[
C_F\geq0.70,\ H_F\leq0.35,\ M_F\geq0.20,\ Q_F\geq0.60,\
10^{-5}\leq\rho_F\leq0.30
\right].
\end{equation}

For each teacher family $T\in\{\mathrm{M},\mathrm{L}\}$, class
probabilities were converted to ET/TC/WT region probabilities,
thresholded at 0.50, and hierarchically corrected to obtain teacher-specific masks $M_r^T$. Inter-teacher agreement for region $r$ was then measured as
\begin{equation}
D_r^{\mathrm{M,L}}
=
\frac{2|M_r^{\mathrm{M}}\cap M_r^{\mathrm{L}}|}
{|M_r^{\mathrm{M}}|+|M_r^{\mathrm{L}}|}.
\end{equation}
ET agreement was accepted when both masks were empty, when the only
nonempty mask had volume below $5\,\mathrm{mm}^3$, or otherwise when $D_{\mathrm{ET}}^{\mathrm{M,L}}\geq0.50$. Denoting this condition by $I_{\mathrm{ET}}$, the final criteria were
\begin{equation}
I_{\mathrm{agree}}
=
\mathbf{1}[D_{\mathrm{WT}}^{\mathrm{M,L}}\geq0.85]\,
\mathbf{1}[D_{\mathrm{TC}}^{\mathrm{M,L}}\geq0.70]\,
I_{\mathrm{ET}},
\qquad
I_{\mathrm{strict}}=I_{\mathrm{conf}}I_{\mathrm{agree}}.
\end{equation}
QC was applied at the case level: failure of any required criterion excluded the entire candidate rather than only the affected region.
The region-fusion weights favored the teacher with stronger observed region-wise predictions: ResEnc-M for ET, equal weighting for TC, and ResEnc-L for WT. QC thresholds were set conservatively to reject low-confidence, high-entropy, low-margin, low-agreement or implausible-coverage predictions while retaining a sufficient unlabeled pool. The relaxed ET criterion accounts for absent or very small enhancing regions. These values were not grid-searched and were fixed across all experiments.
The pre-QC small-component rules were likewise fixed heuristics intended to remove only tiny, weakly supported islands and were not separately optimized.
Accepted masks were converted to mutually exclusive targets by assigning
ET to $M_{\mathrm{ET}}$, NCR/NET to
$M_{\mathrm{TC}}\setminus M_{\mathrm{ET}}$, ED to
$M_{\mathrm{WT}}\setminus M_{\mathrm{TC}}$, and BG otherwise.
Overall, 983 of 1,138 candidates (86.4\%) satisfied
$I_{\mathrm{strict}}$ and were included in every student fold.
Pseudo-labeled cases were used only for student training and were not included in the annotated validation splits.

\subsection{Student Network and Mixed-Supervision Objective}
\label{sec:student-objective}

\paragraph{Backbone and auxiliary supervision.}
The student is a six-stage 3D ResidualEncoderUNet with 32, 64, 128,
256, 320, and 320 channels and 1, 3, 4, 6, 6, and 6 encoder blocks,
respectively~\cite{isensee2024revisited}. Its deep-supervised main
objective \(\mathcal{L}_{\mathrm{seg}}\) equally combines cross-entropy
and soft Dice loss for the four mutually exclusive classes, excluding
background from the Dice term.

A training-only \(1\times1\times1\) auxiliary head predicts independent
ET, TC, and WT regions from the highest-resolution output. With
\(\ell_r=\mathcal{L}_{\mathrm{BCE},r}
+\mathcal{L}_{\mathrm{Dice},r}\), the auxiliary objective is
\begin{equation}
\mathcal{L}_{\mathrm{aux}}
=
0.2\ell_{\mathrm{ET}}
+
0.2\ell_{\mathrm{TC}}
+
0.1\ell_{\mathrm{WT}}.
\label{eq:auxiliary-loss}
\end{equation}
The auxiliary head is used only to compute the training loss and is bypassed by the inference forward path.

\paragraph{Dense expert adapter.}
A residual adapter with \(K\) convolutional experts is inserted at the
final 320-channel encoder skip; the submitted model uses \(K=4\). Each
expert applies group normalization followed by
\(320\!\rightarrow\!80\!\rightarrow\!80\!\rightarrow\!320\)
convolutions with $1^3$, $3^3$, and $1^3$ voxel kernels, respectively.
For bottleneck feature \(\mathbf{z}_i\) of input patch \(i\), a
\(320\!\rightarrow\!80\!\rightarrow\!K\) router produces 
\begin{equation}
\mathbf{g}_i
=
\operatorname{softmax}
\left(
\frac{R(\operatorname{GAP}(\mathbf{z}_i))}{\tau}
\right),
\qquad
\mathbf{z}'_i
=
\mathbf{z}_i
+
\alpha
\sum_{e=1}^{K}
g_{i,e}E_e(\mathbf{z}_i),
\label{eq:dense-expert-adapter}
\end{equation}
where \(\tau=1\), \(\sum_e g_{i,e}=1\), and the trainable residual
scale \(\alpha\) is initialized to \(0.1\). All experts process every
patch, and their outputs are combined by the input-conditioned soft
gates.

To discourage mean gate collapse, we use
\begin{equation}
\bar{g}_e = \frac{1}{B}\sum_{i=1}^{B} g_{i,e},
\qquad
\mathcal{L}_{\mathrm{bal}}
=
\sum_{e=1}^{K}
\left(\bar{g}_e-\frac{1}{K}\right)^2,
\qquad
\lambda_{\mathrm{bal}}=0.001.
\end{equation}

The \(K=4\) dense adapter added 0.93M parameters and remains active at inference.

\paragraph{Mixed-supervision objective.}
Let \(\mathcal{G}\) and \(\mathcal{P}\) denote the annotated and
pseudo-labeled samples in a mini-batch. For
\(S\in\{\mathcal{G},\mathcal{P}\}\), define
\(\mathcal{L}_S=\mathcal{L}_{\mathrm{seg}}^S+
\mathcal{L}_{\mathrm{aux}}^S\), averaged over subset \(S\), with an
empty subset contributing zero. At fine-tuning epoch \(t\), the
objective is
\begin{equation}
\mathcal{L}_t
=
\mathcal{L}_{\mathcal{G}}
+
\lambda_p(t)\mathcal{L}_{\mathcal{P}}
+
\lambda_{\mathrm{bal}}\mathcal{L}_{\mathrm{bal}},
\qquad
\lambda_p(t)
=
\begin{cases}
0.3, & t<50,\\
0.3+0.4(t-50)/100, & 50\leq t<150,\\
0.7, & t\geq150.
\end{cases}
\label{eq:mixed-supervision-objective}
\end{equation}
Thus, both segmentation and auxiliary losses from pseudo-labeled
samples are down-weighted during early fine-tuning.

\section{Experiments}

\subsection{Training and Submission Inference}
\label{sec:training-inference}

\paragraph{Training and implementation.}
Images were resampled to 1-mm isotropic spacing and normalized by masked
case-wise $z$-scores, using third-/first-order image/label interpolation.
Augmentations were rotation ($p=.2$, $\pm30^\circ$), scaling
($p=.2$, .7--1.4), noise ($p=.1$, variance 0--.1), blur
($p=.2$, $\sigma=.5$--1), brightness/contrast
($p=.15$ each, .75--1.25), gamma with/without inversion
($p=.1/.3$, .7--1.5), low-resolution simulation
($p=.25$, .5--1), and per-axis mirroring ($p=.5$).

The ResEnc-M/L teachers were trained only on annotated cases using the
same preprocessing and augmentation pipeline. They used the same five-fold split for 1,000 epochs
(250 training and 50 validation iterations/epoch), patches of
$128\!\times\!160\!\times\!112$/$160\!\times\!192\!\times\!160$, and
batches of 2/3. Both used SGD with Nesterov momentum .99, initial learning
rate .01, weight decay $3\times10^{-5}$, and polynomial decay.

Each student was initialized from the corresponding best ResEnc-M checkpoint;
its new head and adapter were trained for 500 epochs with the same iteration
counts, $128\!\times\!160\!\times\!112$ patches, batch size 2, and foreground
oversampling .5. Student SGD used Nesterov momentum .99, initial learning rate
$5\times10^{-4}$, weight decay $3\times10^{-5}$, polynomial decay, and
gradient clipping at 12.

Annotated and pseudo-labeled cases were pooled and replacement-sampled using
normalized fold-specific ET-aware weights (1.1--2.0), with no fixed source
quota; annotated cases contributed approximately 55.5\% of sampling mass.
Best checkpoints maximized the exponential moving average of mean
non-background validation Dice. Training used one RTX~5090 per fold, CUDA
autocast, and dynamic scaling; checkpoint-recorded active times were
21.4--25.6~h for ResEnc-M teacher folds, 51.7--89.5~h for ResEnc-L teacher
folds, and 10.3--22.0~h for student folds, excluding initialization,
interruptions, and final sliding-window validation.

Pseudo-label generation for all 1,138 cases took 1:54:55/3:12:03
(6.1/10.1 s/case) for ResEnc-M/L five-fold inference and 2:01:16
(6.4 s/case) for fusion and QC, totaling 7:08:14
(22.6 s/case).

\paragraph{Five-fold inference.}
The final ResEnc-M submission used the best checkpoint from each of five
$K=4$ folds. Each applied Gaussian-weighted sliding-window inference using
the training patch size and step size .5, and averaged the eight three-axis
mirror orientations. Fold logits were equally averaged, 
softmax normalized,
and restored to the original space. Four-class probabilities were converted
to ET/TC/WT as in Eq.~\ref{eq:region-probabilities}, thresholded at .50,
hierarchically corrected, and mapped back to mutually exclusive labels.
No connected-component, volume, or morphological post-processing was applied.
Submission-time inference, including the five folds and eight mirror
orientations, took approximately 15.5--15.7 s per case.

\subsection{Evaluation Metrics and Comparison Protocol}
\label{sec:evaluation-protocol}

The validation server reports DSC and NSD for ET, TC, and WT. We compute
unweighted regional means and
\begin{equation}
\overline{\mathrm{DSC}}
=
\frac{1}{3}\sum_{r\in\mathcal{R}}\mathrm{DSC}_r,
\qquad
\overline{\mathrm{NSD}}
=
\frac{1}{3}\sum_{r\in\mathcal{R}}\mathrm{NSD}_r,
\qquad
A_{\mathrm{calc}}
=
\frac{\overline{\mathrm{DSC}}+\overline{\mathrm{NSD}}}{2}.
\label{eq:evaluation-summary}
\end{equation}
Here $A_{\mathrm{calc}}$ is derived from server-reported regional metrics,
not independently reported by the server. Its Panoptica NSD tolerance was
$\tau=.5$, whereas the final ranking uses $\tau=1$; our values therefore
describe the validation server setting.

Quantitative performance within the organizer-released annotated training set
was assessed on the 271-case fold-0 validation split, whereas final performance
was assessed on the official validation server. The auxiliary analysis compares matched five-fold ResEnc-M
configurations without the dense adapter, differing only in auxiliary
supervision. Dense-expert capacity was compared using fold-0 validation server results
with identical pseudo-label, auxiliary, and inference settings; only
$K=4$ was run across all folds.

The heterogeneous ensemble equally averaged class logits before softmax, but
model-family implementations were not fully matched, so this comparison is
descriptive. Local pseudo-label and paired-adapter analyses are exploratory
rather than strictly fold-safe because teacher checkpoints had used annotations
from the corresponding student validation fold.

\section{Results}
\label{sec:results}

\subsection{Auxiliary-Region Supervision and Pseudo-Label Utilization}
\label{sec:auxiliary-supervision}
\paragraph{Auxiliary-region supervision.}
\begin{table}[!htbp]
\centering
\caption{Matched five-fold ResEnc-M validation server results without
the dense adapter, with auxiliary supervision off/on; $\Delta$ is
On minus Off.}
\label{tab:auxiliary-supervision}
\setlength{\tabcolsep}{2.8pt}
\renewcommand{\arraystretch}{0.95}
\small
\begin{tabular}{lccccccccr}
\toprule
& \multicolumn{4}{c}{DSC} & \multicolumn{4}{c}{NSD} & \multirow{2}{*}{$A_{\mathrm{calc}}$} \\
\cmidrule(lr){2-5} \cmidrule(lr){6-9}
Auxiliary & ET & TC & WT & Mean & ET & TC & WT & Mean & \\
\midrule
Off & .787 & .815 & .883 & .828 & .547 & .498 & .488 & .511 & .6695 \\
On & .790 & .815 & .882 & .829 & .550 & .500 & .488 & .513 & .6710 \\
$\Delta$ & +.003 & .000 & -.001 & +.001 & +.003 & +.002 & .000 & +.002 & +.0015 \\
\bottomrule
\end{tabular}
\end{table}

Auxiliary supervision increased $A_{\mathrm{calc}}$ by .0015, with small gains
in ET DSC/NSD and TC NSD but not WT. We therefore interpret it as a modest,
region-dependent regularizer; its branch is bypassed at inference.

\paragraph{Fold-0 pseudo-label utilization.}

\begin{table}[!htbp]
\centering
\caption{Fold-0 validation DSC within the organizer-released annotated training
set (271 cases). All arms use the same ResEnc-M student with the $K = 4$ dense expert adapter; n denotes valid ET cases.}
\label{tab:pseudolabel-utilization}
\begin{tabular}{@{}lcccccc@{}}
\toprule
Setting & Pseudo & QC & ET ($n$) & TC & WT & Mean \\
\midrule
Labeled-only
& 0
& --
& .865 (268)
& .916
& .931
& .904 \\

All-pseudo
& 1{,}138
& No
& .864 (269)
& .919
& .935
& .906 \\

QC-pseudo
& 983
& Strict
& .872 (267)
& .918
& .935
& .908 \\
\bottomrule
\end{tabular}
\end{table}

All-pseudo used all 1,138 candidates after fusion, hierarchy correction,
and the same pre-QC cleanup, whereas QC-pseudo retained only the 983 cases satisfying $I_{\mathrm{strict}}$; all other settings were matched. Following the challenge metric convention, both-empty case-region pairs were excluded from regional means, yielding 268/269/267 valid ET cases for labeled-only/all-pseudo/QC-pseudo, respectively. Mean DSC increased from .904 (labeled-only) to .906 (all-pseudo) and
.908 (QC-pseudo): the unfiltered gain was mainly in TC/WT and the
additional QC gain mainly in ET. Because the differences were small and
the comparison was single-fold and non-fold-safe, this is preliminary
evidence, does not establish a region-specific mechanism, and is not
directly comparable with validation server results.

\paragraph{Strict-QC exclusion audit.}
\begin{table}[!htbp]
\centering
\caption{Non-exclusive failure frequencies among pseudo-label candidates
excluded by strict QC.}
\label{tab:qc_exclusion_audit}
\begin{tabular}{lcc}
\toprule
Failure criterion & Frequency & Percentage (\%) \\
\midrule
\multicolumn{3}{l}{\textit{Nonempty excluded candidates ($n=94$)}}\\
Insufficient high-confidence foreground fraction ($Q_F$)
    & 55/94 & 58.5 \\
High entropy
    & 46/94 & 48.9 \\
Low confidence
    & 37/94 & 39.4 \\
Low margin
    & 25/94 & 26.6 \\
\addlinespace
\multicolumn{3}{l}{\textit{All excluded candidates ($n=155$)}}\\
WT disagreement
    & 94/155 & 60.6 \\
Foreground-coverage range violation ($\rho_F$)
    & 77/155 & 49.7 \\
Empty foreground
    & 61/155 & 39.4 \\
TC disagreement
    & 52/155 & 33.5 \\
ET disagreement
    & 34/155 & 21.9 \\
\bottomrule
\end{tabular}
\end{table}

Table~\ref{tab:qc_exclusion_audit} reports non-exclusive failures;
$C_F$, $H_F$, $M_F$, and $Q_F$ use the 94 nonempty exclusions, and
the remaining criteria use all 155 exclusions.

For retained/excluded nonempty cases ($n=983/94$), median (Q1--Q3)
values were $C_F$: .9823 (.9682--.9869)/.7435 (.5805--.8684),
$H_F$: .0302 (.0226--.0514)/.3427 (.2000--.4814),
$M_F$: .9650 (.9374--.9741)/.4873 (.1640--.7397), and
$Q_F$: .9706 (.9459--.9789)/.4894 (.0833--.7369); these measures
were undefined for the other 61 excluded cases.

Across all retained/excluded cases ($n=983/155$), corresponding values
were $\rho_F$: $2.74\times10^{-3}$
($5.23\times10^{-4}$--$9.27\times10^{-3}$)/$1.14\times10^{-5}$
(0--$9.07\times10^{-5}$), $D_{\mathrm{WT}}^{\mathrm{M,L}}$:
.9845 (.9710--.9906)/.8027 (.4220--1.0000),
$D_{\mathrm{TC}}^{\mathrm{M,L}}$: .9903
(.9819--.9937)/.9506 (.5326--1.0000), and pre-QC ET volume:
10,448 (3,084.5--28,582.5)/5.0 (0--317.5) mm$^3$.

These data characterize filtering, not accuracy or artifact/cohort bias,
because ground truth, image-quality labels, and case metadata were unavailable.

\subsection{Dense Expert Capacity}
\label{sec:expert-capacity}

\begin{table}[!htbp]
\centering
\caption{Validation server results from fold-0 ResEnc-M models with
different dense expert capacities.}
\label{tab:expert-capacity}
\setlength{\tabcolsep}{2.8pt}
\renewcommand{\arraystretch}{0.95}
\small
\begin{tabular}{lccccccccr}
\toprule
& \multicolumn{4}{c}{DSC}
& \multicolumn{4}{c}{NSD}
& \multirow{2}{*}{\(A_{\mathrm{calc}}\)} \\
\cmidrule(lr){2-5}
\cmidrule(lr){6-9}
Adapter & ET & TC & WT & Mean & ET & TC & WT & Mean & \\
\midrule
No adapter
& .781 & .809 & .880 & .823
& .537 & .492 & .477 & .502
& .6625 \\
\(K=2\)
& .771 & .808 & .879 & .819
& .531 & .490 & .483 & .501
& .6600 \\
\(K=3\)
& .787 & .806 & .878 & .824
& .546 & .492 & .481 & .506
& .6650 \\
\(K=4\)
& \textbf{.798} & .803 & .879 & \textbf{.827}
& \textbf{.554} & .492 & .479 & \textbf{.508}
& \textbf{.6675} \\
\(K=5\)
& .769 & .805 & .880 & .818
& .535 & \textbf{.494} & .477 & .502
& .6600 \\
\bottomrule
\end{tabular}
\end{table}
All configurations differed only in adapter capacity. Relative to no
adapter, $K=4$ achieved the highest $A_{\mathrm{calc}}$ (.6675) and
increased ET DSC/NSD by .017, while TC/WT DSC changed by
$-.006/-.001$; $K=2$ and $K=5$ did not exceed the control. We therefore
selected $K=4$ on fold~0 without claiming universal optimality.

A paired local analysis on the same 271 fold-0 cases compared $K=4$ with
the auxiliary-matched no-adapter control. For ET/TC/WT, median
$\Delta$Dice (IQR) was .00143 ($-.00160$--.00527), $-.00047$
($-.00219$--.00265), and .00015 ($-.00123$--.00186);
improved/tied/worsened proportions were 62.4/3.7/33.9\,
53.1/1.1/45.8\, and 53.9/.4/45.8\%; and Wilcoxon signed-rank
$p$-values were $2.01\times10^{-6}$, .297, and .106. Thus, ET showed a
statistically detectable but small paired shift, whereas TC and WT did
not. Greater ET heterogeneity and optimization complexity are possible
explanations, but expert specialization was not tested.

\subsection{System-Level Model-Family and Ensemble Comparison}
\label{sec:model-family-comparison}
\begin{table}[!htbp]
\centering
\caption{Five-fold validation server system-level results.}
\label{tab:model-family-comparison}
\setlength{\tabcolsep}{2.8pt}
\renewcommand{\arraystretch}{0.95}
\small
\begin{tabular}{lccccccccr}
\toprule
& \multicolumn{4}{c}{DSC}
& \multicolumn{4}{c}{NSD}
& \multirow{2}{*}{\(A_{\mathrm{calc}}\)} \\
\cmidrule(lr){2-5}
\cmidrule(lr){6-9}
Model & ET & TC & WT & Mean & ET & TC & WT & Mean & \\
\midrule
ResEnc-M
& .797 & .823 & .881 & \textbf{.834}
& .547 & .501 & .481 & .510
& \textbf{.6720} \\
MedNeXt
& .791 & .804 & .874 & .823
& .556 & .500 & .480 & .512
& .6675 \\
BI-SegMamba
& .781 & .800 & .875 & .819
& .543 & .490 & .480 & .504
& .6615 \\
Three-model ensemble
& .797 & .805 & .879 & .827
& .559 & .500 & .488 & \textbf{.516}
& .6715 \\
\bottomrule
\end{tabular}
\end{table}

ResEnc-M achieved the best mean DSC (.834) and $A_{\mathrm{calc}}$
(.6720), whereas the heterogeneous ensemble achieved the best mean NSD
(.516); because implementations were not fully matched, this comparison
is descriptive.

\section{Discussion}

Component effects were modest and region-dependent: fold-0 analyses showed
small gains from pseudo-label augmentation and strict QC, modest auxiliary
effects, and a clear paired adapter shift only for ET; heterogeneous ensembling
increased mean NSD but reduced mean DSC. Strict QC was evaluated as a whole,
without criterion-level ablation. Excluded candidates had weaker
confidence-related measures, coverage, and WT/TC inter-teacher agreement,
higher entropy, and smaller pre-QC ET volumes. Without ground truth,
image-quality labels, or case-level metadata, these descriptive differences
cannot establish prediction error, artifact effects, or phenotype- or
cohort-specific bias.

Internal component analyses remain exploratory because teacher
checkpoints had used annotations from the corresponding student
validation fold, allowing indirect influence through pseudo-labels.
The pseudo-label comparison was single-fold, $K$ was selected on the
same fold, and neither the non-monotonic capacity pattern nor the
gate-balancing objective establishes functional expert specialization.
Future work will generate pseudo-labels for student fold $k$ only from
teachers excluding fold $k$ and conduct fully matched paired analyses
across folds.

\section{Conclusion}

We presented a mixed-supervision system combining quality-controlled pseudo-labeling, nested-region supervision, and dense expert adaptation. The final validation server result was mean DSC 0.834, mean NSD 0.510, and $A_{\mathrm{calc}} = 0.6720$. Component effects were region-dependent; the auxiliary branch is bypassed at inference, whereas the dense adapter remains active. Future work will use fold-safe teachers and fully matched paired analyses.

\begin{credits}
\subsubsection{\ackname}
Data used in this publication were obtained as part of the Challenge project
through Synapse ID (syn74274097). This research was supported in part by a
grant of the Korea Health Technology R\&D Project through the Korea Health
Industry Development Institute (KHIDI), funded by the Ministry of Health \&
Welfare, Republic of Korea (grant number: RS-2025-02222326).

\subsubsection{\discintname}
The authors have no competing interests to declare that are relevant to the
content of this article.
\end{credits}

\bibliographystyle{splncs04}
\bibliography{references}

\end{document}